\RequirePackage{fix-cm}
\documentclass[pdflatex,sn-mathphys-num]{sn-jnl}

\usepackage{graphicx}%
\usepackage{multirow}%
\usepackage{amsmath,amssymb,amsfonts}%
\usepackage{amsthm}%
\usepackage[scr=rsfs]{mathalfa}
\usepackage[title]{appendix}%
\usepackage{xcolor}%
\usepackage{textcomp}%
\usepackage{manyfoot}%
\usepackage{booktabs}%
\usepackage{algorithm}%
\usepackage{algorithmicx}%
\usepackage{algpseudocode}%
\usepackage{listings}%

\usepackage[utf8]{inputenc}
\usepackage[T1]{fontenc}
\usepackage{mathtools}
\usepackage{siunitx}
\usepackage{bm}
\usepackage{gensymb}
\usepackage{placeins}
\usepackage{hyperref}

\usepackage[utf8]{inputenc}
\usepackage{amsmath}
\usepackage{amsfonts}
\usepackage{graphicx}
\usepackage[export]{adjustbox}
\usepackage[svgnames,dvipsnames]{xcolor}

\usepackage{subcaption}
\usepackage{tikz}
\usepackage{gensymb}
\usetikzlibrary{patterns, decorations.pathreplacing}
\usetikzlibrary{patterns, arrows.meta, calc, decorations.pathreplacing}
\definecolor{waterblue}{RGB}{180, 220, 255} 

\usepackage{natbib} 
\usepackage{amssymb}
\usepackage{amsthm}
\usepackage{graphicx}
\usepackage{epstopdf, epsfig}
\usepackage{bm}

\usepackage{placeins}
\usepackage{pgfplots}
\usepgfplotslibrary{fillbetween}
\pgfplotsset{compat=1.15}
\newlength\figH
\newlength\figW
\usepackage{bm}
\usepackage{overpic} 
\usepackage{tabularx}  

\usepackage{array}     
\usepackage{mathtools}  
\usepackage{adjustbox} 
\pgfplotsset{compat=newest}

\definecolor{cadmiumorange}{rgb}{0.93, 0.53, 0.18}

\usepackage{hyphenat}

\newcommand{\We}{\mathrm{We}}
\renewcommand{\Re}{\mathrm{Re}} 
\newcommand{\Oh}{\mathrm{Oh}}

\theoremstyle{thmstyleone}%
\theoremstyle{thmstyletwo}%

\theoremstyle{thmstylethree}%

\begin{document}

\title[Compound drop impacts]{Compound drop impacts on substrates of different temperatures}

\author[1,2]{\fnm{Benjamin} \sur{Gorin}}
\author[2,3]{\fnm{Mete} \sur{Abbot}}\email{mete.abbot@gmail.com}
\author[2]{\fnm{Daniel} \sur{Bonn}}\email{d.bonn@uva.nl}
\author[1]{\fnm{Hamid} \sur{Kellay}}\email{hamid.kellay@u-bordeaux.fr}

\affil[1]{Laboratoire Ondes et Matiere d’Aquitaine, University of Bordeaux, 33400 Talence, France}
\affil[2]{Institute of Physics, University of Amsterdam, Science Park 904, 1098 XH Amsterdam, the Netherlands}
\affil[3]{Department of Aerospace Engineering, Faculty of Engineering, Dokuz Eyl\"ul University, 35390 Izmir, Turkey}


\abstract{In this paper, we report the impacts of isothermal compound drops for various combinations of viscous fluids. We extend an existing parameter-free model based on kinematic closure of drop impact to predict the maximum spreading diameter of both the core and the shell of compound drops. Then, we report impacts of compound drops on a cold surface.
By using two liquids with different melting/freezing temperatures, it is possible to freeze either one of the two liquids or both and doing so, different patterns of frozen drops are obtained.}


\maketitle

\section{Introduction}

Understanding how a droplet impacts and spreads over a surface is crucial for many applications and has been extensively investigated, as reported in numerous studies \cite{worthington_xxviii_1877,chandra_collision_1991, bartolo_retraction_2005,bartolo_dynamics_2007, bonn_wetting_2009, laan_maximum_2014, lee_universal_2016, gordillo_theory_2019, gorin_universal_2022,abbot2026kinematic}. Drop impacts occur in various situations, such as irrigation in crop fields using sprays, coating processes in industry, or when aerosols are used and inhaled for drug delivery \cite{glasser_tuning_2019,xie_delamination_2021,lohse_fundamental_2022, sijs_how_2021}.

A more complex situation occurs when the so-called compound drops impact a surface \cite{johnson_fluid_1985,kan_hydrodynamics_1998,chiu_experiment_2005,chen_impact_2007,terwagne_bouncing_2010,liu_maximal_2018,liu_ejecting_2019,blanken_rebound_2020,blanken_impact_2021}.
A compound drop refers to a drop made of two or more different immiscible phases whether they are solid, liquid or air phases. The general structure of such drops is a core of one phase surrounded by a shell of the other phase. These compound drops can be of great interest for different applications : One can think about the modification of the rheology of the assembly of drops, the transport and deposition of solid particles or drug delivery, the creation of miniaturised microfluidic device or even cryopreservation.

For isothermal impacts of compound drops, a recent study by Blanken et.al \cite{blanken_rebound_2020} reported that a partial rebound of the core drop occurs during the impact. This phenomenon known as a self-lubrication mechanism comes from the presence of the outer layer of the compound drop. They also showed that retraction and jetting of the core drop strongly depends on the relative position of the core drop. Above a critical height and if the density of the liquid core drop is superior to the shell liquid, the free fall time of the drop becomes sufficiently long so that the core drop migrates, due to gravity, to the bottom of the compound drop before impacting the surface. In that case, the core drop breaks the lubricated layer and spreads directly over the substrate.

A partial jetting also happens in hollow droplet impacts, as first shown by Gulyaev \cite{gulyaev_hollow_2012,nasiri_hollow_2021}. The entrapped air bubble within the droplet deforms the droplet at the impact, leading to the formation of a counter jet. 
For a hot surface, the compound drop rebounds after the impact due to Leidenfrost effect. Chiu et.al \cite{chiu_experiment_2005} showed that the presence of a core drop made of a different liquid affects the rebound and the breakup of the compound drop at the impact.

Although the maximum spreading diameter of pure liquid drop impacts has been widely investigated \cite{madejski_droplets_1983,roisman_normal_2002,laan_maximum_2014,lee_drop_2010,josserand_drop_2016,gordillo_theory_2019,abbot2026kinematic}, including emulsion drops \cite{piskunov_effect_2021,piskunov2025spreading,abbot2025elucidating,damak2022dynamics}, only very few studies focused on the understanding of the maximum spreading diameter of compound drop impacts, where a single drop is surrounded by another drop of immiscible liquid. \citet{liu_maximal_2018} have studied numerically the impact and spreading of liquid compound drops and found that the maximum spreading diameter depends on the volume ratio and the interfacial tension of the core drop with its liquid shell. 

In this paper, we report the impacts of isothermal compound drops for various combinations of viscous fluids; the maximum spreading diameter of the core and the shell drop is compared to simple drop impacts. Then, we report impacts of compound drops on a cold surface.
By using two liquids with different melting/freezing temperatures, it is possible to freeze either one of the two liquids or both and doing so, different patterns of frozen drops are obtained. 


\section{Impact and spreading of an isothermal compound drop}\label{sec3}

\subsection{Water-in-oil compound drop}

\begin{figure}[!ht]
     \centering
     \includegraphics[width=1\textwidth]{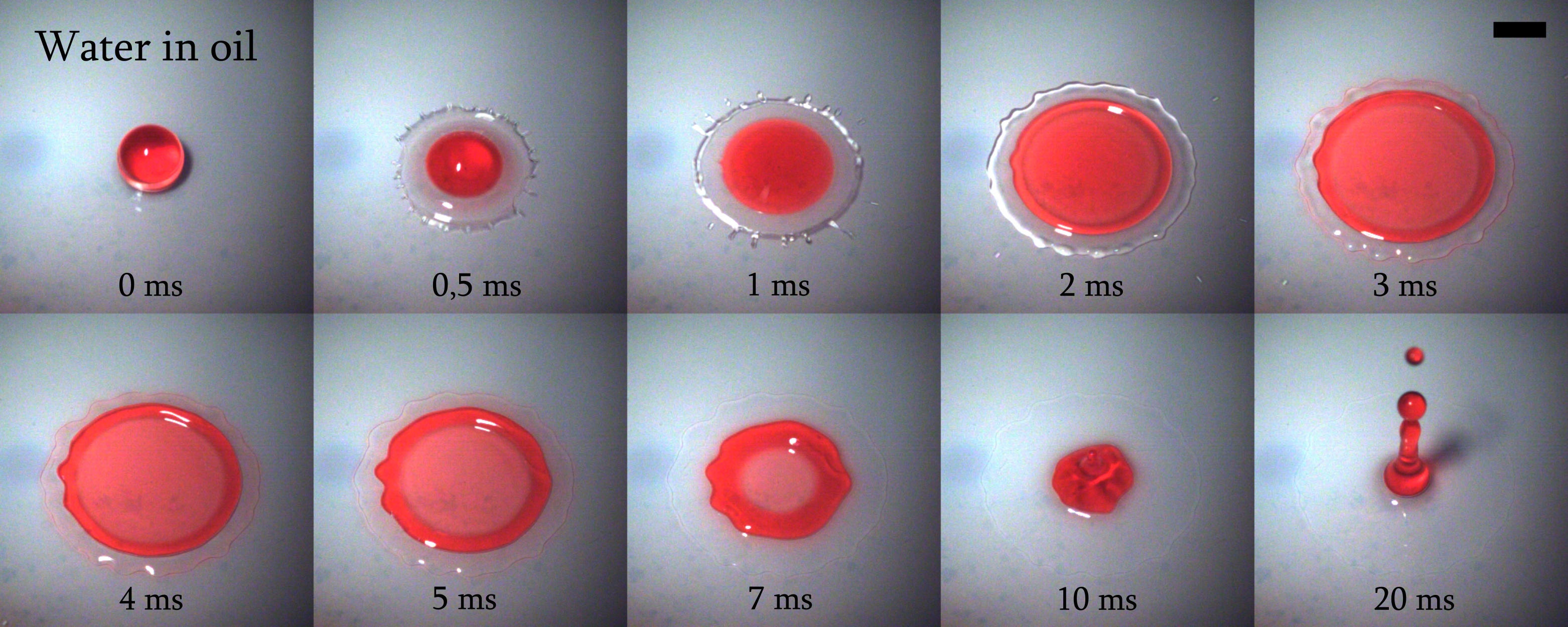}
        \caption{Timelapse of the impact of a water-in-oil compound drop on a hydrophilic glass surface. Red dye is added to the water phase. The impact velocity is $V_0=2 \ m.s^{-1}$, the drop size is $D_0=2.56 \ mm$ and the volume ratio of the compound drop is $\alpha \approx 0.5$. Scale bar: 2mm.}
        \label{fig: Timelapse water-in-oil compound isothermal}
\end{figure}

We first examine the impact and spreading of a water-in-oil compound drop, as illustrated in Figure \ref{fig: Timelapse water-in-oil compound isothermal}. 
Upon impact, the compound drop spreads fast with outer liquid or shell liquid spreading first. The outer oil shell wets the substrate and no retraction is observed. The water core on the other hand which spreads over a thin layer of oil retracts rapidly and gives rise to the formation of a jet as on a hydrophobic surface. This behaviour has been reported by \citet{blanken_rebound_2020} who showed that the partial rebound of the water core (on a hydrophilic substrate) comes from a self-lubrication mechanism: upon impact, the water drop spreads over a thin layer of oil which comes from the surrounding shell drop. Subsequently, the water drop recoils, resulting in the formation of a jet. The jetting behavior of the core drop depends on the liquid properties (surface tension, viscosity, density), the impact speed $V_0$ and the volume ratio $\alpha$ between the core and the shell.

\begin{figure}[!ht]
     \centering
        \includegraphics[width=1\textwidth]{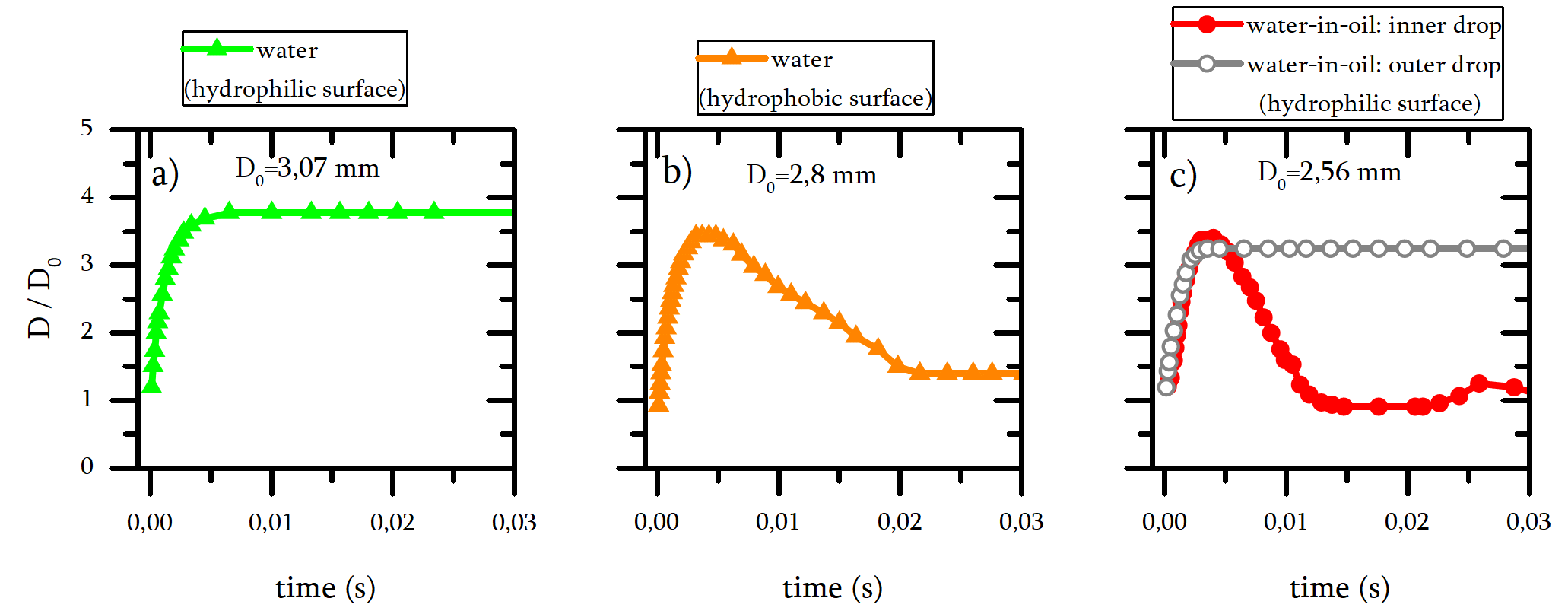}
        \caption{Rescaled spreading diameters of : a) a drop of pure water impacting a hydrophilic glass surface, b) a water drop impacting a hydrophobic glass surface and c) a water-in-oil compound drop impacting a hydrophilic glass surface. Drop sizes are given in each graphs. The impact velocity is kept constant for each drops: $V_0 = 2 \ m.s^{-1}$ and  the volume ratio of the compound drop is $\alpha \approx 0.5$.}
        \label{fig: Spreading comparison water-in-oil}
\end{figure}

In order to understand how the spreading of a compound drop differs from that of a simple drop, we compare on Figure \ref{fig: Spreading comparison water-in-oil} the spreading diameter dynamics of pure water drops impacting on either hydrophobic (Figure \ref{fig: Spreading comparison water-in-oil}a) or hydrophilic surfaces (see Figure \ref{fig: Spreading comparison water-in-oil}b) to the one of a compound drop (see Figure \ref{fig: Spreading comparison water-in-oil}c). 
As expected, the water drop spreads and wets the hydrophilic glass surface while it spreads then retracts on the hydrophobic surface. On the other hand, the compound drop impacting the hydrophilic glass, the water core spreads and retracts like a simple water drop on a hydrophobic surface: the presence of a surrounding liquid oil layer drastically affects the spreading of the water drop. The oil shell however spreads in a similar fashion as a simple drop. 

\subsection{Oil-in-water compound drop}\label{subsec2}

The drop configuration depends on the difference between the surface tensions of the two liquid phases so to achieve an oil drop surrounded by a water shell, it is necessary to reduce the surface tension of the water-air interface. 
The surface tension of water air interface is reduced by dissolving solid Dioctyl sulfosuccinate sodium salt (an ionic surfactant commonly known as AOT) in pure water up to a concentration $C_{\mathrm{AOT}}=20 \ mM$. 
As a result, the surface tension of water with the surfactant $\gamma_{\mathrm{AOT}} \approx 26 \ mN.m^{-1}$ \cite{sijs_how_2021} becomes lower than the surface tension of hexadecane $\gamma_{oil}=27 \ mN.m^{-1}$. Doing so, we can create an oil drop encapsulated in water.

\begin{figure}[!ht]
     \centering
     \includegraphics[width=1\textwidth]{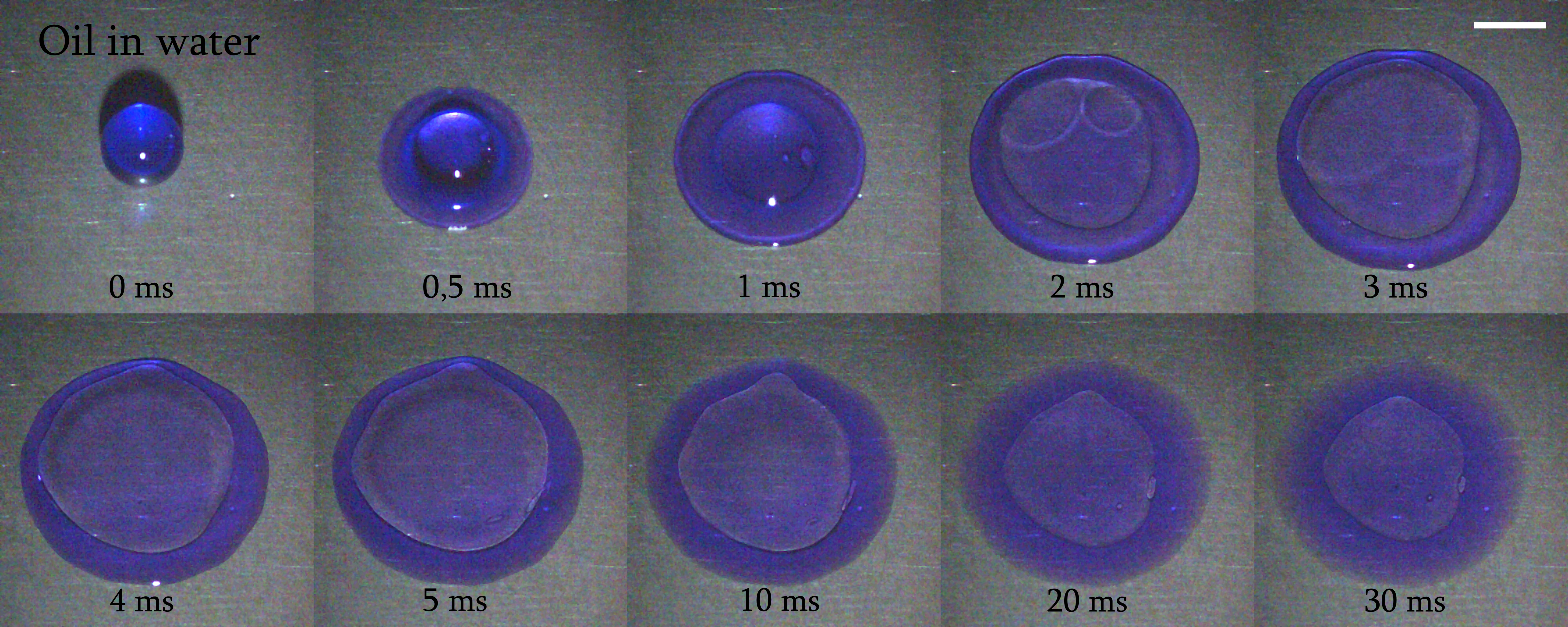}
        \caption{Timelapse of the impact of an oil-in-water compound drop on an aluminium surface. Dye is added to the water phase. The impact velocity is $V_0=2 \ m.s^{-1}$, the drop size is $D_0=2.36 \ mm$ and the volume ratio of the compound drop is $\alpha \approx 0.5$. Scale bar: 2 mm.}
        \label{fig: Timelapse oil-in-water compound isothermal}
\end{figure}

\begin{figure}[!ht]
     \centering
        \includegraphics[width=1\textwidth]{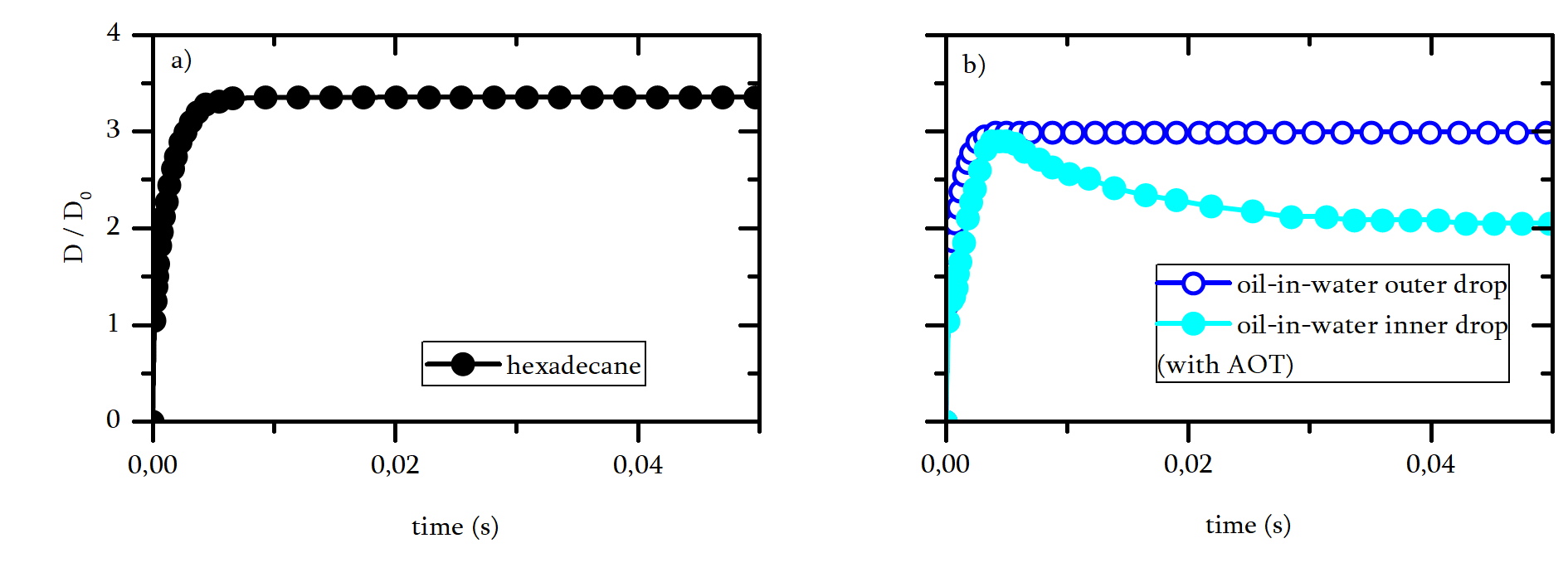}
        \caption{Rescaled spreading diameters of : a) a hexadecane oil drop impacting an aluminium surface, b) an oil-in-water compound drop impacting a hydrophilic aluminium surface. The impact velocity is kept constant for each drops: $V_0 = 2 \ m.s^{-1}$ and  the volume ratio of the compound drop is $\alpha \approx 0.5$.}
        \label{fig: spreading oil-in-water}
\end{figure}

A timelapse of this type of compound drop impacting an aluminium
surface is shown in Figure \ref{fig: Timelapse oil-in-water compound isothermal}. The water, dyed in blue, spreads similarly to the simple water droplet on a hydrophilic surface showing no retraction at longer times. On the contrary, the transparent oil core drop, spreads over the water drop and then retracts until it reaches an equilibrium diameter. We also see that the thin layer of water above the oil core drop breaks at two different spots during the spreading between 1 and 2 ms after the impact. The water film covering the oil core drop becomes thinner as the drop spreads and at some point, very small perturbations (air flow, hydrodynamic instabilities, air bubbles trapped at the moment droplet detaches from the needle.) are enough to break the top water-air liquid film.
The spreading dynamics of both the oil core and the water shell drop are shown in Figure \ref{fig: spreading oil-in-water}.
Compared to the water-in-oil compound drop, here the retraction of the oil core drop is much slower. Therefore, we have two different examples of compound drops showing two distinct spreading behaviours.


\section{Maximum spreading diameter of isothermal compound drop impacts}

\subsection{Maximum spreading models}

We now examine quantitatively the maximum spreading diameter $D_{\mathrm{max}}$ of such isothermal compound drops upon impact. The maximum spreading diameter of drop impacts has been the subject of numerous studies, resulting in the proposal of different scaling laws for various regimes. The most well-known are the inertio-viscous regime  scaling of $\sim \Re^{1/5}$ and the inertio-capillary regime scaling of $\We^{1/2}$. These two regimes typically require fitting with adjustable prefactors. Furthermore, the two regimes were bridged using a Pade approximant proposed by \citet{laan_maximum_2014} in the form initially proposed by \citet{eggers_drop_2010} as:

\begin{equation}
    \frac{D_{\mathrm{max}}}{D_0} \sim \Re^{1/5} f(P)
    \label{eq4: Dmax scaling law compound drop}
\end{equation}

with $f(P) = \frac{\sqrt{P}}{A+\sqrt{P}}$, $D_0$ the initial drop diameter and $A=1.24$ is a fitting parameter \cite{laan_maximum_2014}. The variable $P$, known as the impact parameter, is defined as $P=\We\Re^{-2/5}$, where $\We$ and $\Re$ are the Weber and Reynolds numbers, respectively. 
For simple drop impacts, the Weber and Reynolds numbers are defined as $\Re=\rho V_0 D_0 / \eta$ and $\We=\rho {V_0}^2 D_0 / \gamma$.

\citet{laan_maximum_2014} equation was later shown to significantly deviate from experiments for high viscosity droplets or low impact parameter $P$ \cite{abbot2026kinematic,liu2025maximum,sanjay2025unifying}. The deviation was initially proposed to stem from zero-velocity impact finite wetting, and were proposed to be fixed by subtracting a fitted value of $\beta_0$ \cite{lee_modeling_2016,de_goede_effect_2018}. Recent kinematic closure by \citet{abbot2026kinematic} fixed this problem by proposing energy balance based equations for spreading time $t_s$ and spreading average velocity $V_s$ that satisfy the kinematic closure of $D_{\mathrm{max}}= t_s V_s$, leading to the unified parameter-free maximum spreading equation:
\begin{equation}
    \beta_{\text{max}} = \frac{\sqrt{\We + 12} }{1 + \Oh^{1/5}(\We +12)^{2/5}}. 
\end{equation}
This equation is strictly for the maximum contact diameter ratio $\beta_c$. For impacts where $\beta_c \ge 2$, contact diameter and equatorial diameter are close. However, for $\beta_c < 2$, \citet{abbot2026kinematic} showed that the geometrical correction $\beta_w = 0.25 \beta_c^2 +1$, where $\beta_w$ is the equatorial diameter ratio to initial diameter $D_0$, approximates the equatorial diameter ratio within error significantly less than typical experimental error. 

For this work, we analyze the results using the \citet{abbot2026kinematic} framework to obtain a parameter free model for both the maximum spreading diameter of the outer and inner droplets. In the appendix, we compare with \citet{laan_maximum_2014} using fitted effective impact parameters which altogether lead to satisfactory results with only 3 fitting parameters.

\subsection{Outer Maximum Diameter $D_{\mathrm{max, out}}$}

Here, we define the Weber and Reynolds numbers for both phases. 

For the outer drop, $\Re_{\mathrm{out}}$ and $\We_{\mathrm{out}}$ are defined as :

\begin{equation}
    \Re_{\mathrm{out}}=\frac{\rho_{\mathrm{eff}} V_0 D_{\mathrm{0,out}}}{\eta_{\mathrm{out}}} \quad \quad \quad \We_{\mathrm{out}}=\frac{\rho_{\mathrm{eff}} {V_0}^2 D_{\mathrm{0,out}}}{\gamma_{\mathrm{out}}}
    \label{eq: Re and We outer drop}
\end{equation}

The viscosity and the surface tension are the same as those of the outer layer. The drop size corresponds to the compound drop size. However, the density is given by the effective density which takes into account the density of the core drop. The effective density is $\rho_{\mathrm{eff}}=\left( \alpha \rho_{\mathrm{in}} + \rho_{\mathrm{out}}\left( 1-\alpha\right)\right)$. Ohnesorge number is then $\Oh_{\mathrm{out}} = \sqrt{\We_{\mathrm{out}}}/\Re_{\mathrm{out}}$. The extended \citet{abbot2026kinematic} model for the spreading ratio of the outer phase $\beta_{\max,\mathrm{out}}$ becomes:
\begin{equation}
\beta_{\max,\mathrm{out}} \equiv \frac{D_{\max,\mathrm{out}}}{D_0} \approx \frac{\sqrt{\We_{\mathrm{out}} + 12}}{1 + \left[ \Oh_{\mathrm{out}} (\We_{\mathrm{out}} + 12)^2 \right]^{1/5}},
\label{eq:closure_out}
\end{equation}
which predicts the results with error less than experimental error, as shown in Figure~\ref{fig:outer_diameter}.

\begin{figure*}[!t]
    \centering
    \makebox[\textwidth][c]{%
        \includegraphics[width=1.2\textwidth]{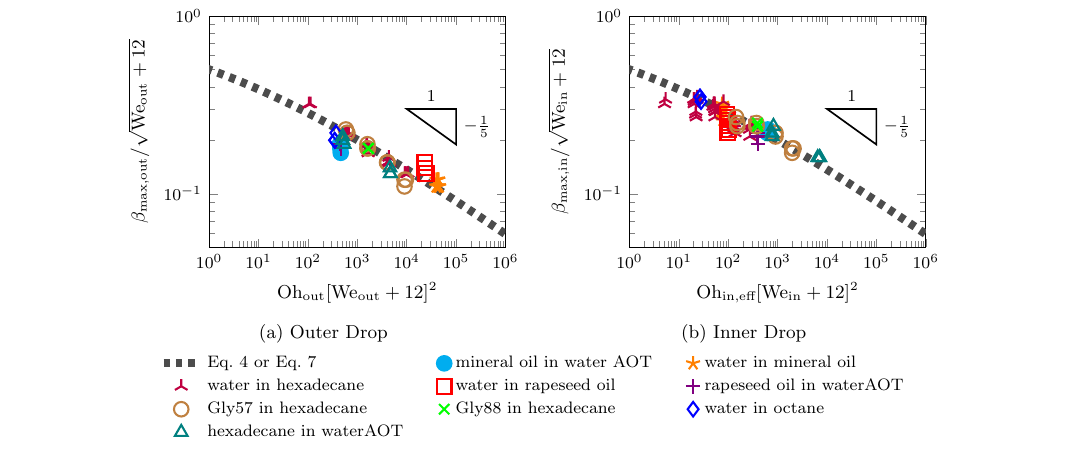}%
    }
    \caption{Dimensionless maximum spreading ratio collapsed against the unified damping parameter for (a) the outer shell against Eq.~\ref{eq:closure_out} and (b) the inner core against Eq.~\ref{eq:closure_in}. Across all fluid pairs and viscosity ratios, both phases collapse onto the theoretical prediction without adjustable parameters.}
    \label{fig:outer_inner_diameter}
\end{figure*}

\subsection{Inner Maximum Diameter $D_{\mathrm{max, in}}$}

The initial equivalent diameter of the inner core is determined from volume conservation 
as $D_{0,\mathrm{in}} = \alpha^{1/3} D_{0,\mathrm{out}}$. In contrast to the outer shell, 
which experiences a no-slip boundary condition at the solid substrate, the inner core spreads inside the other liquid, therefore experiencing an effective viscosity resulting from both liquids resisting the flow. 

To obtain a leading-order estimate of the viscous dissipation without empirical prefactors, we approximate the lamella in the lubrication limit as a two-layer Couette-type shear flow. Enforcing the continuity of shear stress across the liquid--liquid interface 
($\tau \approx \eta_{\mathrm{out}} \dot{\gamma}_{\mathrm{out}} \approx \eta_{\mathrm{in}} \dot{\gamma}_{\mathrm{in}}$), 
with the layer thicknesses scaling with their respective volume fractions 
($h_{\mathrm{out}} \propto 1 - \alpha$ and $h_{\mathrm{in}} \propto \alpha$), yields an effective 
series dissipation viscosity:
\begin{equation}
    \eta_{\mathrm{in,eff}} \approx \left( \frac{1 - \alpha}{\eta_{\mathrm{out}}} + \frac{\alpha}{\eta_{\mathrm{in}}} \right)^{-1} 
    = \frac{\eta_{\mathrm{out}} \eta_{\mathrm{in}}}{\alpha \eta_{\mathrm{out}} + (1 - \alpha) \eta_{\mathrm{in}}}.
    \label{eq:eta_in_eff}
\end{equation}
For this work, $\alpha=0.5$ is fixed across all experiments. While this idealized relation neglects several secondary factors, such as lateral rim drag, transient boundary-layer 
development, and contact line friction, it captures the dominant bulk dissipation and converges to the correct limits, i.e. $\alpha \to1$, then $\eta_{\mathrm{in,eff}} \to \eta_{\mathrm{in}}$ and $\alpha \to 0$, then $\eta_{\mathrm{in,eff}} \to \eta_{\mathrm{out}}$.

The effective surface tension is taken as $\sigma_{\max} = \max(\gamma_{\mathrm{in}}, \gamma_{\mathrm{out}})$, 
where $\gamma_{\mathrm{in}}$ is the liquid--liquid interfacial tension \cite{wu_dynamic_1999,zeppieri_interfacial_2001} 
and $\gamma_{\mathrm{out}}$ is the outer liquid--air surface tension. With the core density $\rho_{\mathrm{in}}$, 
the inner Weber and effective Reynolds numbers are defined as:
\begin{equation}
     \Re_{\mathrm{in,eff}} = \frac{\rho_{\mathrm{in}} V_0 D_{0,\mathrm{in}}}{\eta_{\mathrm{in,eff}}},
     \quad 
     \We_{\mathrm{in}} = \frac{\rho_{\mathrm{in}} V_0^2 D_{0,\mathrm{in}}}{\sigma_{\max}}.
\end{equation}
The effective Ohnesorge number follows as $\Oh_{\mathrm{in,eff}} = \sqrt{\We_{\mathrm{in}}} / \Re_{\mathrm{in,eff}} = \eta_{\mathrm{in,eff}} / \sqrt{\rho_{\mathrm{in}} \sigma_{\max} D_{0,\mathrm{in}}}$.

Applying the extended \citet{abbot2026kinematic} kinematic closure to the inner droplet yields:
\begin{equation}
    \beta_{\max,\mathrm{in}} \equiv \frac{D_{\max,\mathrm{in}}}{D_{0,\mathrm{in}}} \approx \frac{\sqrt{\We_{\mathrm{in}} + 12}}{1 + \left[ \Oh_{\mathrm{in,eff}} (\We_{\mathrm{in}} + 12)^2 \right]^{1/5}}.
    \label{eq:closure_in}
\end{equation}
As shown in Figure~\ref{fig:inner_diameter}, this parameter-free closure collapses the maximum spreading ratio 
of the inner core across a wide range of viscosity ratios, capturing the experimental data 
within typical experimental uncertainty without freely adjustable parameters.

\section{Impact of a compound drop on a cold surface}

Having examined the effect of mismatch in viscosity on the spreading dynamics of compound drops, we now turn to the effects of the substrate and the substrate temperature in particular. Consider the case where the drop impacts a surface cooled below the freezing temperature of the liquid. The spreading diameter of the drop is generally reduced \cite{herbaut_liquid_2019, herbaut_criterion_2020, thievenaz_freezing-damped_2020, kant_fast-freezing_2020, koldeweij_initial_2020, gorin_droplet_2022} : $D_{\mathrm{end}}<D_{\mathrm{max}}$.
What happens when the drop is composed of two immiscible liquids? How does the heterogeneity of the drop influence the spreading and the freezing dynamics compared to a simple drop impact? We are not aware of studies elucidating such questions and which we address in the following.

Here we examine the impact and spreading behavior of a compound drop impacting a cold solid surface. The two liquids composing the drop have different freezing temperatures and in general, different thermal properties. In order to limit the number of parameters, we only consider compound drops composed of water and Hexadecane. While the freezing of water occurs at $0\degree C$, that of Hexadecane occurs at $18\degree C$ leaving a large range of temperatures where only one of the liquids freezes.

\begin{figure}[!ht]
    \centering
    \includegraphics[width=1\textwidth]{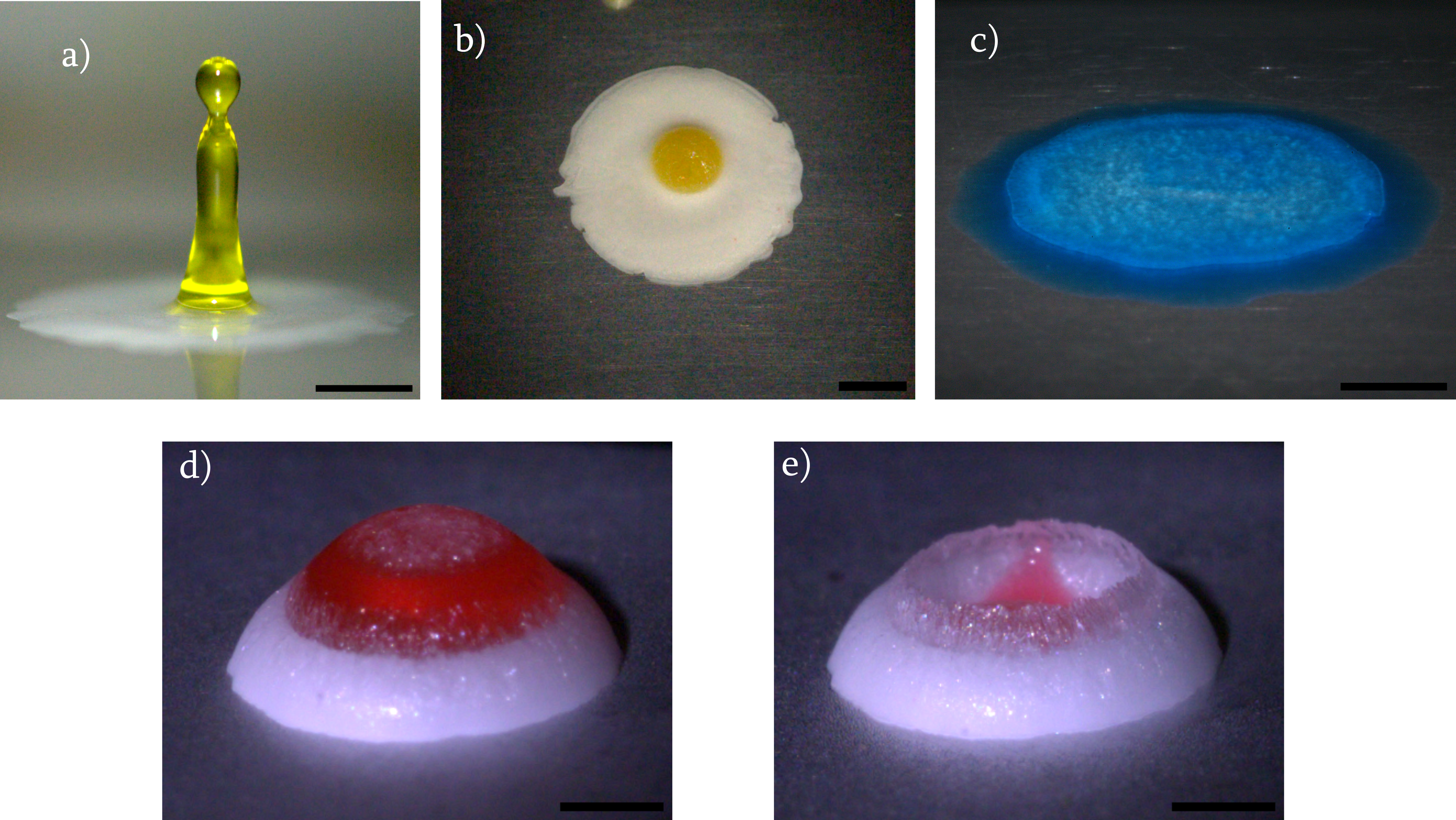}
    \caption{Different behaviours of a compound drop impact on a cold aluminium surface. The compound drop is made of a dyed water phase and an oil phase (hexadecane). The two phases have different melting temperatures (see Table \ref{tab:liquid_properties}) allowing to freeze either one or both liquids. a) Water-in-oil compound drop $13 \ ms$ after the impact: The oil phase freezes while the water drop spreads, retracts and jets. $V_0=2 \ m.s^{-1}$ and $T_s=10 \degree C$ (see Movie 1  in SI). b) Water-in-oil compound drop $5s$ after the impact: the oil phase is frozen while the water core remains liquid in a supercooled state. $V_0=2 \ m.s^{-1}$ and $T_s=-15 \degree C$. c) An oil-in-water compound drop impact. The water phase dyed in blue, remains liquid while the oil phase freezes and floats on the water drop. $V_0=1,5 \ m.s^{-1}$ and $T_s=7 \degree C$. d) Oil-in-water compound drop $100 \ ms$ after the impact: both water (dyed in red) and hexadecane phase freezes. $V_0=2 \ m.s^{-1}$ and $T_s=-15 \degree C$. d) A water-in-oil compound drop deposited on a cold surface ($T_s=0\ \degree\mathrm{C}$), imaged once the oil phase is completely frozen; the liquid water core (dyed in red) is trapped within the frozen oil crater. e) The liquid water is removed using a clean tissue to reveal the crater structure of the frozen oil shell. Scale bars: $2\ \mathrm{mm}$.}
    \label{fig:frozen-compounddrops}
\end{figure}

Different scenarios can be thus be anticipated depending on the temperature of the surface: First, there is the isothermal case when the surface temperature is at room temperature, higher than the melting point of both the  oil and water phases. This is the case examined above. Next, for the range $0<T_s<18 \degree C$, only the oil phase, made of Hexadecane, may freeze during the impact. Below $0 \degree C$, both Hexadecane and water are likely to freeze although water may remain in a super-cooled liquid state depending on the conditions of the experiment. Each such situation will engender specific spreading dynamics which we consider below for both water-in-oil and oil-in-water compound drops. In Figure~\ref{fig:frozen-compounddrops}, we illustrate a few cases of compound drops impacting a cold surface. One can note that different cases ensue such as the freezing of the outer shell only (a), the freezing of both the shell and the core (b), or the freezing of the inner liquid only( c). In addition, different shapes can be obtained such as a frozen fried egg like pattern (b), a floating island (c), or a crater like shape obtained at low impact velocities and where the outer fluid forms a crater around the inner liquid which remains fluid. 

\subsection{Impact of a water-in-oil compound drops on a cold surface}

\begin{figure}[!ht]
    \centering
    \includegraphics[width=0.9\textwidth]{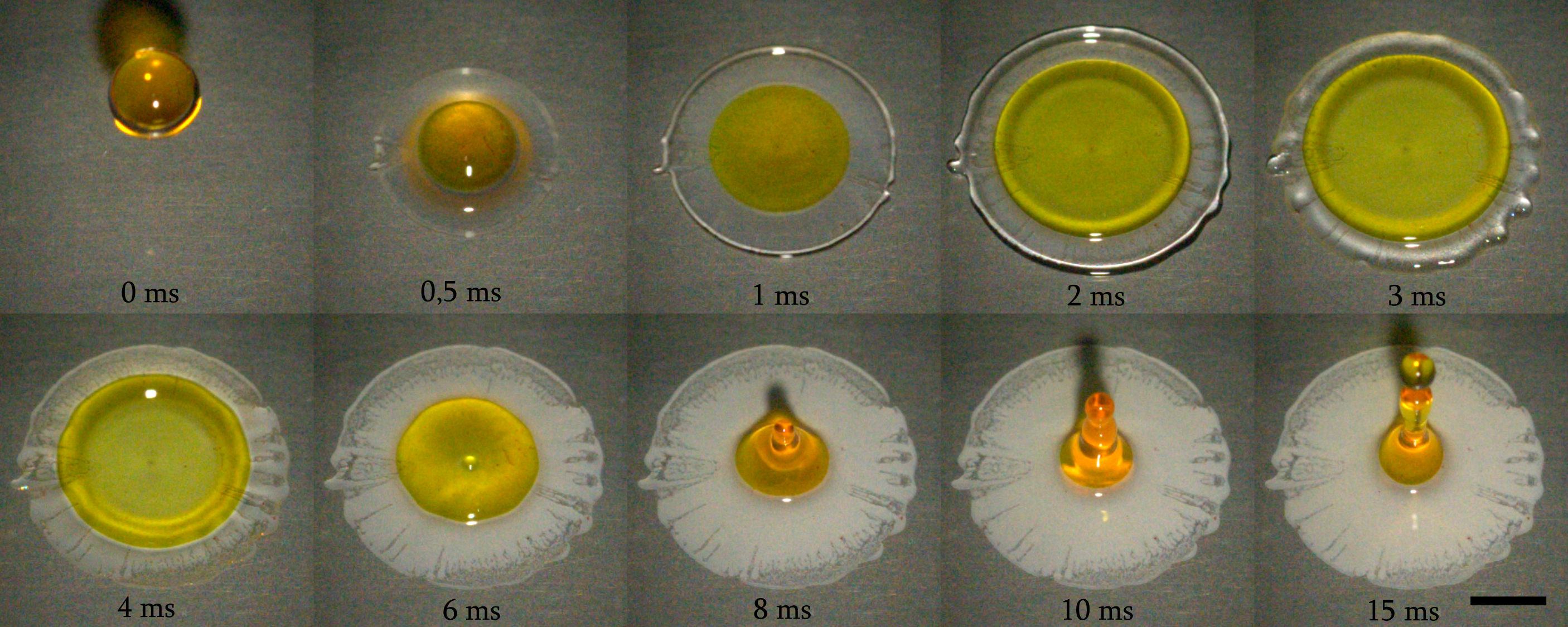}
    \caption{Timelapse of a water-in-oil compound drop impacting a cold surface. Hexadecane freezes during the spreading of the drop but does not affect significantly the spreading of the core which spreads, retracts and jets. The surface is made of aluminium, cooled down to $T_s=-15 \degree C$. Impact speed is $V_0=2 \ m.s^{-1}$.}
    \label{fig: timelapse water-in-oil cold}
\end{figure}

We first examine the impact of compound drops made of a water core and an oil shell. The oil phase made of Hexadecane has a freezing point of $T_{f,oil} = 18 \degree C$ which is higher than the freezing point of water $T_{f,water} = 0 \degree C$.
To investigate the influence of freezing on the spreading dynamics, several impacts of compound drops on cold aluminium surfaces are recorded for different surface temperatures and two different impact velocities. The experimental setup used is the same as for the isothermal impacts of compound drops (see Figure \ref{fig4: compound drop intro}).
A timelapse of the impact of a water-in-oil compound drop is shown on figure \ref{fig: timelapse water-in-oil cold}. 

\begin{figure}[!ht]
    \centering
    \includegraphics[width=0.82\textwidth]{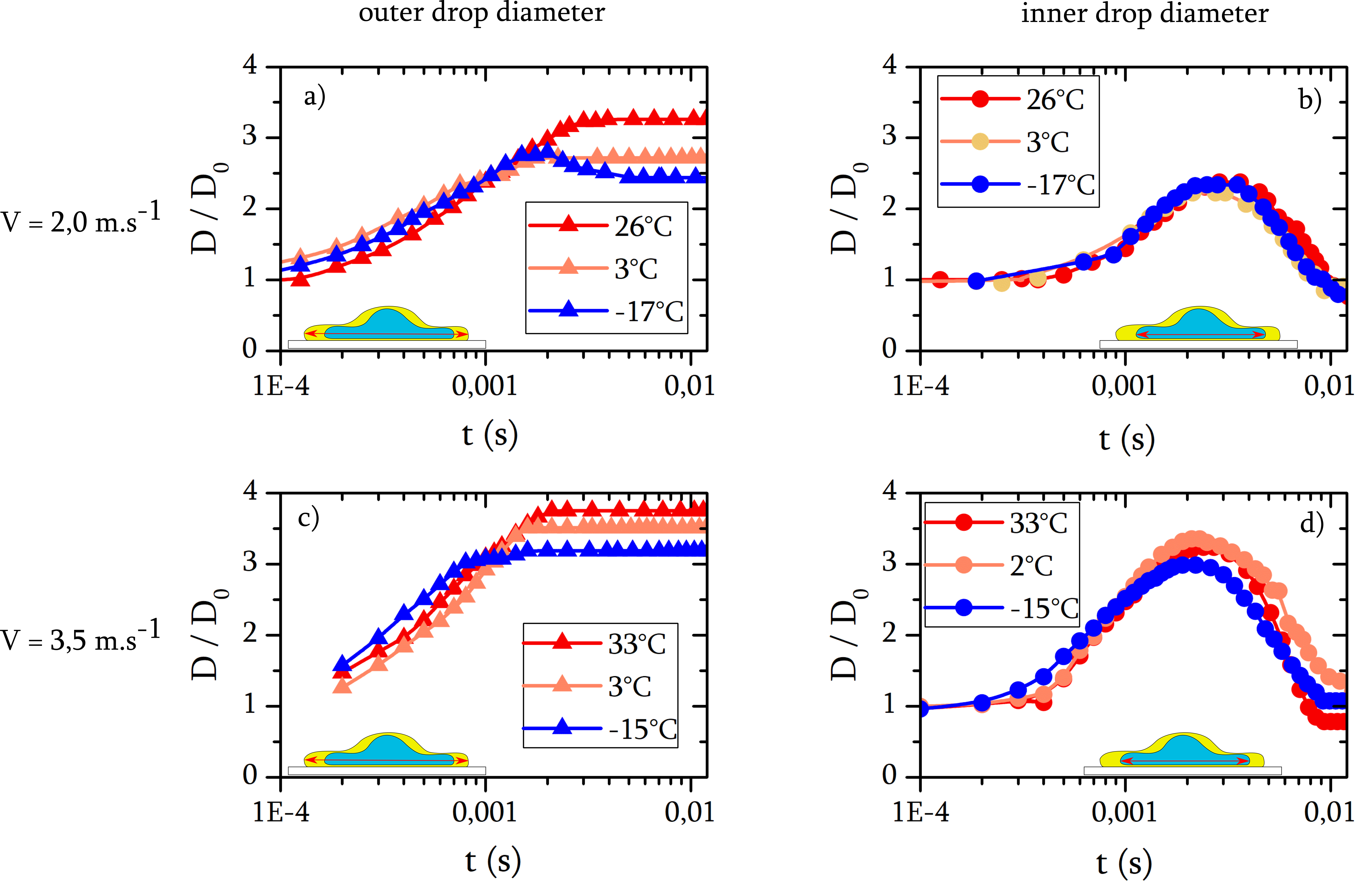}
    \caption{Spreading dynamics of water-in-oil compound drops impacting at two different velocities and for different surface temperature. a-) Spreading dynamics of the core and the shell drop for $V_0=2 \ m.s^{-1}$. c-d) Spreading dynamics of the core and the shell drop for $V_0=3.5 \ m.s^{-1}$}
    \label{fig: Spreading dynamics water-in-oil cold}
\end{figure}

When such a drop impacts a surface cooled below the freezing temperature of the oil, the oil layer in contact with the surface freezes, resulting in the arrest of the contact line and a reduced final spreading diameter (see Figure \ref{fig: timelapse water-in-oil cold}).
The temperature of the surface significantly affects both the spreading and the final shape of the compound drop.  This observation is similar to the case of simple drop impacts on cold surfaces which has been reported previously \cite{madejski_solidification_1976,thievenaz_solidification_2019,thievenaz_retraction_2020,thievenaz_freezing-damped_2020}. The water core drop then spreads over this thin layer of frozen oil and in most cases, exhibits no discernible effect of the surface temperature compared to the isothermal impacts (see Figure \ref{fig: Spreading dynamics water-in-oil cold}) and that for the two considered impact velocities: The presence of an oil shell around a water drop strongly delays the freezing of the latter.

Note that for the highest impact velocity $V_0=3.5 \ m.s^{-1}$ and the lower surface temperature $T_s=-15 \degree C$, the core drop spreads up to a maximum diameter slightly lower than the isothermal case. In this particular example, the freezing of the outer layer limits the spreading of the water core.
The corresponding spreading dynamics are given in Figure \ref{fig: Spreading dynamics water-in-oil cold}. Note the large recoil of the water core drop as it spreads on a frozen oil layer which modifies the spreading of the liquid water. 
Compared to the isothermal case, the main difference is a small reduction of the maximum spreading diameter of the outer layer. However compared to the bare water drop, the presence of the outer shell delays the freezing of the water core and allows for a large recoil of the drop.  

\subsection{Impact of an oil-in-water compound drop}

\begin{figure}[!ht]
    \centering
    \includegraphics[width=\textwidth]{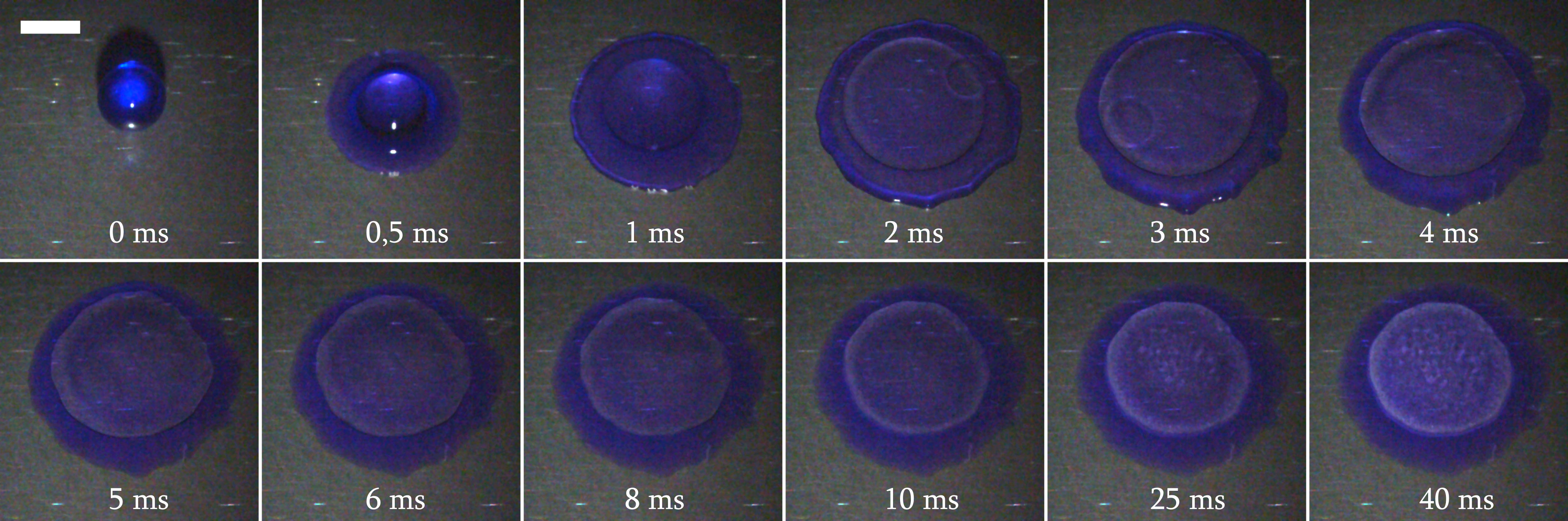}
    \caption{Impact and spreading of an oil-in-water compound drop impacting an cold aluminium surface ($T_s=5 \degree C$) at $V_0=2 \ m.s^{-1}$. The water phase is dyed. Drop size $D_0=2.44 \ mm$ and Scale bar: $2mm$. }
    \label{fig:timelapse oil-in-water cold}
\end{figure}

We now look at the impact of oil-in-water compound drops which are obtained by reducing the surface tension of water with AOT as done for isothermal impacts.
As a consequence, the freezing temperature of the core phase is now higher than that of the shell phase. This means that the oil core drop can be frozen before freezing the water shell (see movie 2 in SI).
A timelapse of the impact of such a drop is shown on Figure \ref{fig:timelapse oil-in-water cold}.
During the impact, the water drop spreads and does not retract because of the very low surface tension of the water solution. Regarding the oil core, it spreads, then retracts and finally freezes. Once the oil is completely frozen, it floats as a thin sheet on the liquid water film. The compound drop thus allows for making thin solid sheets of material at least for moderate impact velocities. 

\begin{figure}[!ht]
    \centering
    \includegraphics[width=0.82\textwidth]{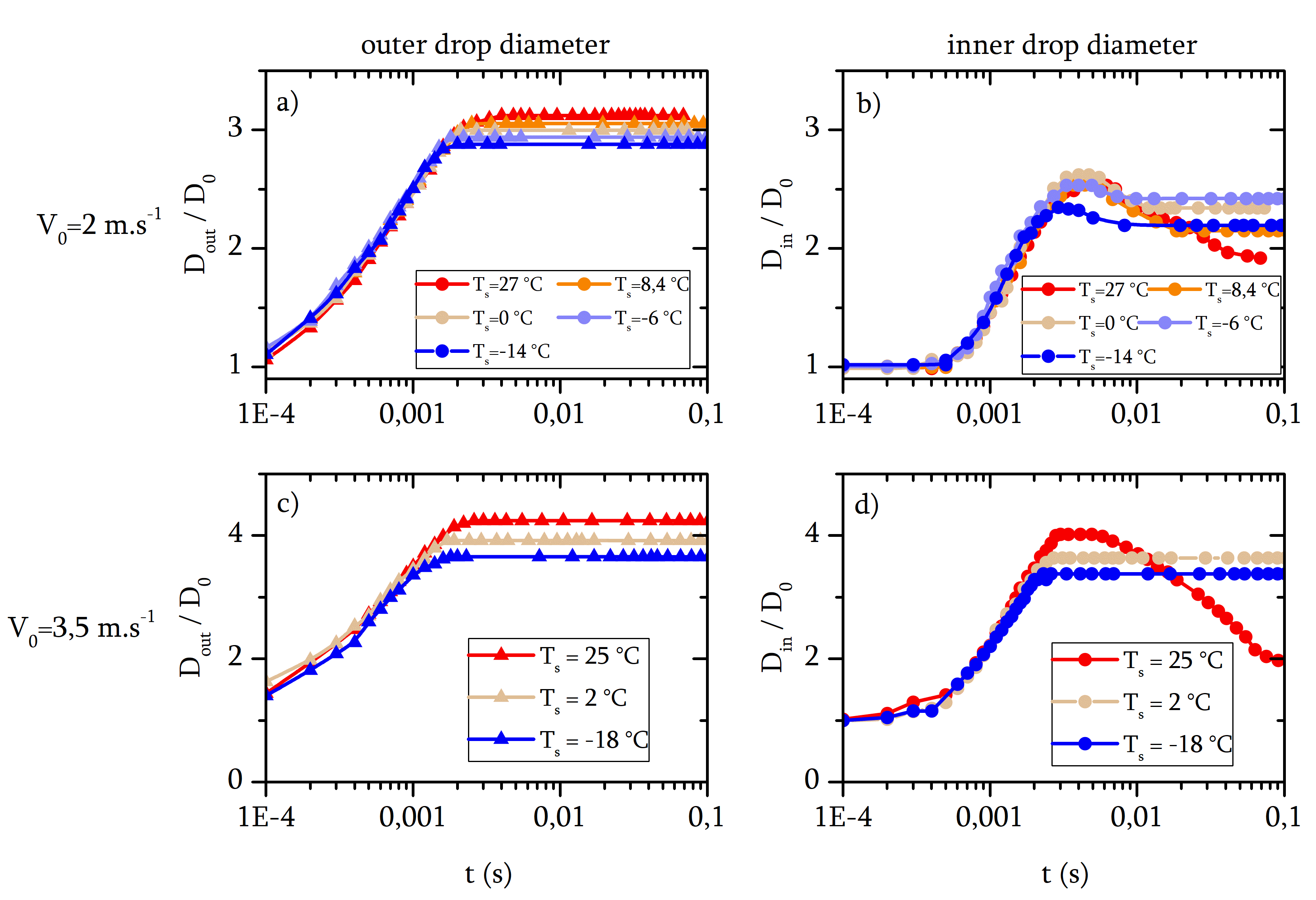}
    \caption{Spreading dynamics of oil-in-water compound drops impacting at two different velocities and for different surface temperature. a-b) Spreading dynamics of the core and the shell drop for $V_0=2 \ m.s^{-1}$. c-d) Spreading dynamics of the core and the shell drop for $V_0=3.5 \ m.s^{-1}$.}
    \label{fig4: spreading on cold surface oil-in-water}
\end{figure}

The dynamics of the spreading diameter for both the oil core and water shell for two different impact velocities and different surface temperatures are illustrated in Figure \ref{fig4: spreading on cold surface oil-in-water} .
For both velocities in the isothermal case, the oil core drop represented in red (in Figure \ref{fig4: spreading on cold surface oil-in-water}b and d) spreads over a thin layer of water and slowly retracts after reaching the maximum spreading diameter. Compared to the water-in-oil compound drops, the retraction of the oil core is considerably slower. This is primarily due to the smaller surface tension of the oil/water interface. On the other hand, the water shell represented in red in Figure \ref{fig4: spreading on cold surface oil-in-water}a and c spreads until it reaches a maximum spreading diameter.

Next, the surface is cooled. For example in Figure \ref{fig4: spreading on cold surface oil-in-water}a and b, at $T_s=0 \ \degree C$, the surface temperature is set below the freezing point of oil which induces the freezing of the oil core. However, the freezing of the oil is delayed due to the spreading over a thin layer of water that acts as a thermal insulator. This can be appreciated through the presence of a maximum in the spreading dynamics before settling on a constant but smaller final diameter due to freezing. Note that the final reduced diameter of the core on a cold surface is greater than the final diameter of the oil core observed in the case of an isothermal drop impact as the recoil is inhibited by the freezing of the liquid. The water phase on the other hand remains liquid and does not freeze, with a spreading dynamics close to the isothermal case.

For the same velocity and a surface temperature of $T_s=-14 \ \degree C$ well below the freezing point of both water and oil, we observe on figure \ref{fig4: spreading on cold surface oil-in-water}a that the maximum spreading diameter of the water shell (blue line) decreases slightly. The spreading of the contact line is arrested due to the freezing, as previously reported for simple drops on cold surfaces \cite{thievenaz_solidification_2019, gorin_droplet_2022}.
The oil core drop on the other hand, represented on Figure \ref{fig4: spreading on cold surface oil-in-water}b, freezes earlier during the retraction. Reducing the surface temperature increases the cooling rate of the oil core resulting in a shorter time of arrest of the spreading of the core drop. This is observed in Figures \ref{fig4: spreading on cold surface oil-in-water}b where the red curve goes below the other curves at lower temperatures indicating a larger final diameter at lower temperatures: freezing eliminates the recoil and thus makes for larger diameters.

For a higher impact speed represented in Figure \ref{fig4: spreading on cold surface oil-in-water}c, the maximum spreading diameter of the shell drop decreases with the temperature below the freezing point of water. The spreading of the oil core drop represented on Figure \ref{fig4: spreading on cold surface oil-in-water}d is also arrested for temperatures below the freezing point of oil.

However, there is a significant difference between the two considered impact velocities : the time of the arrest of the core drop is much shorter for the higher impact velocity: For a cold surface temperature and an impact velocity $V_0=2 \ m.s^{-1}$ (see $T_s=-14 \degree C$ on Figure \ref{fig4: spreading on cold surface oil-in-water}d) the oil core begins to freeze approximately 8ms after the impact.
However, for the higher impact velocity of $V_0=3.5 \ m.s^{-1}$ and a cold surface temperature at $T_s=-18 \degree C$, the oil core starts to freeze around 3ms after the impact, which is close to the freezing time of the water shell drop (see figure \ref{fig4: spreading on cold surface oil-in-water}c,d).
Therefore, the time of the freezing of the oil core drop can be reduced by decreasing the surface temperature or by increasing the impact velocity.

Thus when an oil drop is confined within a water drop, the freezing of the oil core may be delayed with the delay time depending on both the surface temperature and the impact velocity. This is due to the presence of a water layer between the cold surface and the core oil drop, acting as an insulator. 
It is plausible that the thickness of the water layer between the core and the surface depends on the impact speed. An increase in the impact speed results in a larger maximum spreading diameter with a thinner water layer between the core drop and the surface. Consequently, the cooling of the core drop occurs at a faster rate.

\subsection{Final spreading diameter of compound drops}

\begin{figure}[!ht]
    \centering
    \includegraphics[width=1\textwidth]{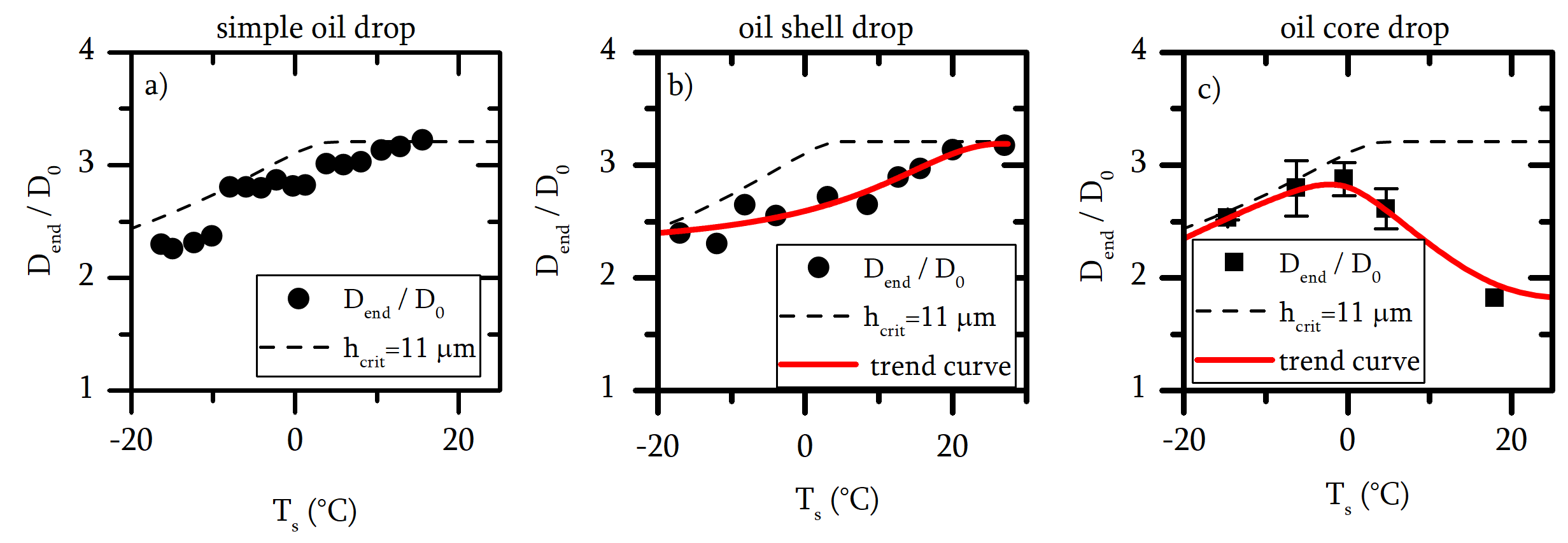}
    \caption{Normalized final spreading diameters $\frac{D_{\mathrm{end}}}{D_{0}}$ of the oil phase for compounds drops impacts on a cold surface. (a) simple oil drop, (b) oil shell of a water-in-oil drop, and (c) oil core of an oil-in-water drop. }
    \label{fig:Dmax cold}
\end{figure}

To better understand how the surface temperature affects the spreading of the oil phase, we report in Figure \ref{fig:Dmax cold} the final diameter of the oil phase in compound drop impacts. 
We first show in Figure \ref{fig:Dmax cold}a the maximum spreading diameter of simple Hexadecane drops impacting a cold surface. As previously reported, the maximum spreading diameter of the oil drop is reduced for temperatures lower than the freezing point of oil. Additionally, we superimpose a model first introduced by Gorin et.al \cite{gorin_droplet_2022} that predicts the arrest diameter of drops impacting on a cold surface. The model is based on a simple criterion: It assumes that a layer of frozen hexadecane forms at the interface of the liquid drop and the solid and that the spreading of the drop is arrested when this layer reaches a critical thickness $h_{crit}$. In the case of hexadecane drops, they found $h_{crit} \approx 11 \mu m$. This model gives a good trend in the case of simple drops as shown in Figure \ref{fig:Dmax cold}a.
On Figure \ref{fig:Dmax cold}b and \ref{fig:Dmax cold}c, we report separately the maximum spreading diameter of the oil phase for the two configurations of compound drops, along with the model of simple hexadecane drops.

In the first configuration of a water core drop encapsulated in a shell drop  made of oil (see Figure \ref{fig:Dmax cold}b), the final diameter of the oil drop decreases as the temperature of the impacted surface decreases. The model slightly overestimates the reduced diameter of the oil shell. The maximum diameter is reduced, for intermediate temperatures, by roughly 10 to 15 \% compared to the simple drops. This is possibly due to the thin layer aspect of the spreading shell. The presence of an inner water drop mildly affects the freezing of the oil shell.

In the configuration of an oil-in-water compound drop, the trend is different as seen on figure \ref{fig:Dmax cold}c. First and because the oil drop retracts after spreading on a water film, the final drop diameter is smaller than for a simple drop. By cooling the surface, the final spreading diameter $D_{\mathrm{end}}$ of the oil core drop increases: the core drop freezes during the retraction phase making for larger spreading diameters. Further and below a certain temperature, decreasing the surface temperature leads to a decrease of the spreading diameter: the core drop starts to freeze during the spreading phase and the presence of the outer water phase is no longer very important. The major difference due to the presence of the water phase comes at intermediate temperatures due to the delay in freezing of the oil allowing it to retract and freeze with smaller diameters (see Figure \ref{fig4: spreading on cold surface oil-in-water}).

\section{Conclusion and perspectives}

We have provided an overview of the diverse patterns and dynamics associated with the impacts of compound drops on a flat surface maintained at different temperatures. Our investigation focuses on measuring the spreading dynamics of these complex drops, as well as on the influence of the liquid properties and the surface  temperature.
The spreading dynamics of the compound drops is governed by the Abbot–Bonn kinematic closure \cite{abbot2026kinematic}, with the inner drop dissipation set by the harmonic series viscosity resulting from interfacial shear continuity.

On cold surfaces, for a water-in-oil compound drop, the oil component freezes upon impact, while the water core spreads similarly to the isothermal case at first before retracting and forming a jet. This behavior can be attributed to the presence of the oil layer, which acts as a thermal insulator modifying both the surface temperature and its wettability .

On the other hand, in the case of an oil-in-water compound drop, the oil freezing is delayed due to the thermal insulation provided by the water layer. However, when the impact velocity is increased, freezing occurs faster. The compound drop spreads more at higher impact velocities, leading to a thinner water layer between the surface and the oil core, consequently increasing the freezing rate.

In summary, our study demonstrates that altering surface temperature, impact speed, and drop configuration yields a diverse range of spreading drop behaviors. Understanding and controlling the spreading of compound drops can be of great interest in various applications, including microencapsulation, emulsion formation, or microfluidics. By manipulating interfacial properties and external conditions, patterns engendered by impacting drops can be designed with specific structures and functionalities.

\section{Methods}

\begin{figure}[ht!]
    \centering
    \includegraphics[width=1\textwidth]{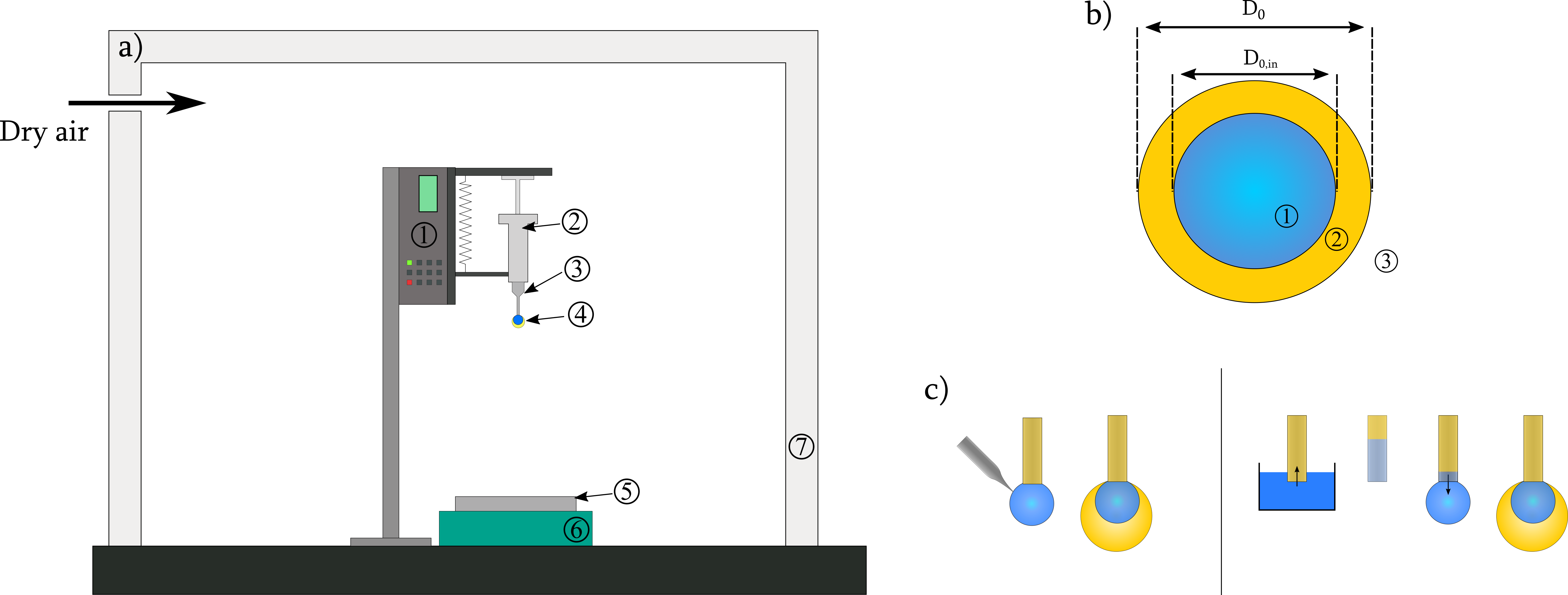}
    \caption{ a) Schematic of the experimental setup. 1: syringe pump, 2: glass syringe, 3: teflon needle, 4: compound drop, 5: aluminium block, 6: cooling block, 7: transparent plastic box. b) Illustration of a two liquid phases compound drop. We define the core phase diameter as $D_{\mathrm{0,in}}$ and the shell drop diameter as $D_0$. c) Illustration of the two different techniques used to produce a compound drop : a) Core drop deposition and b) Suction method. }
    \label{fig4: compound drop intro}
\end{figure}

The compound drops used in this study are made of two concentric drops made of non-miscible liquids: for example, a drop of water encapsulated in a drop of oil, as illustrated in Figure \ref{fig4: compound drop intro} a.

The stability of the compound drop depends on the surface energy of the different interfaces and the stable configuration is the one that minimizes the total surface energy of the compound drop. By tuning the surface tensions of the different phases, it is possible to change the configuration of the compound drop \cite{johnson_fluid_1985}. Here, we restrict our system to a core drop encapsulated in a shell drop.

The making of the compound drop is done by two different methods, the suction method or the deposition method both illustrated in Figure \ref{fig4: compound drop intro} c. The making of the compound drops is important on the configuration of the compound drop :

\begin{itemize}
    \item Coaxial method: The use of a coaxial needle with an inner and outer diameter allow to first produce the shell drop from the outer needle and then the core drop from the inner needle. Compound drops produced from the coaxial method have the core situated at the top of the drop when it is released from the needle.
    
    \item Injection method: a pendant drop is formed at the tip of a needle. Then, the core drop is injected inside the pendant drop using a pipette. If the density of the core is higher than the pendant drop, it sinks down to the bottom of the pendant drop. The geometry of the compound drop when it detaches from the needle is thus different from the coaxial method, with a core drop closer to the bottom of the drop. 
    
\end{itemize}

These two techniques are relatively easy to use. In our experiments, we tested three other techniques to produce a compound drop :

\begin{itemize}

    \item Needle proximity: Here, two needles are put close to each other and inject both liquids at the same time, However, the release of the drop is often affected due to the presence of two needles at the pinch-off.
    
    \item Core drop deposition: Another way to produce a compound drop is to deposit the core drop on the tip of a needle containing the liquid of the shell drop (i.e, the liquid with the lowest surface tension). Then the liquid is pushed out from the needle to form the shell drop around the core. While the shell drop is being produced, the core drop remains at the tip of the needle until the drop detaches. By doing so, the core drop is situated at the top of the compound drop when it is released. 
    
    \item Suction method: This is the technique we mainly use in our experiments. Here, a needle contains the liquid of the shell drop. Then, the needle is immersed in an Eppendorf containing the liquid of the core drop and the desired volume of the core drop is suctioned. After that, the Eppendorf is removed and we push the liquid out of the needle : the core drop will form first, followed by the formation of the shell drop.
\end{itemize}

The first one consists in the deposition of the desired volume of the shell drop at the tip of the needle. The needle contains the other liquid phase. After the deposition, the syringe pump is activated: the inner liquid phase is pushed out, forming the core drop surrounded by the liquid shell drop. The second method consists in absorbing a desired volume of the liquid core drop in the needle. Then the liquid is pushed out and the core drop is first formed, followed by the shell drop which covers the latter. The compound drop detaches from the needle when its weight overcomes the capillary forces acting at the tip of the needle.

\hspace{3mm}

In order to study the effects of viscosity, surface tension and the freezing of compound drops, different combinations of liquids are used.
The properties of these liquids are given in Tables \ref{tab:liquid_properties} and \ref{table4: compound drops}.

The experimental setup for the impacts of the compound drops, illustrated in Figure \ref{fig4: compound drop intro} a, is as follows: a glass syringe (Fortuna Optima glass syringes from Sigma-Aldrich) with a teflon needle is fixed on a syringe pump (Harvard Appartus PHD 2000) and positioned vertically at a specific height. The glass syringe contains the liquid of the shell drop while the core drop is produced at the tip of the needle using the deposition method. The volume of the core drop is chosen to maintain a volume ratio of $\alpha \approx 0.5$. The sizes of the compound drops range from $D_0=2.2$ to $3 \ mm$. The impact velocity $V_0$ is deduced from the release height ($V_0=\sqrt{2gh})$.

Two different surfaces are used: a glass coverslip or an aluminium block. Both surfaces are put in contact with a metal cooling block connected to a cooling system allowing precise control of the surface temperature. The surface is enclosed in a small transparent plastic box where dry air can be injected in order to control the relative humidity of the chamber ($RH<5 \%$) and prevent frost formation on the surface for low cooling temperatures. A fast camera is placed vertically, slightly tilted in order to record the impact from above. 

\begin{table}[ht]
    \centering
    \begin{tabular}{|c|c|c|c|c|c|c|}

 \multicolumn{7}{c}{Liquid phase} \\ 
     \hline
      $-$ & $T_f$  & $\rho$ & $c_p$ & $k$ & $e$ & $L$ \\ \hline
 \hline
        Water & 0 & 1000 & 4217 & 0.555 & 1529.8 & 334000 \\
        Hexadecane & 18 & 767 & 2220 & 0.147 & 504.5 & 236000 \\ \hline \addlinespace
    \end{tabular}
        
    \begin{tabular}{|c|c|c|c|c|}
        \multicolumn{5}{c}{Solid phase} \\
             \hline
        - & $\rho$ & $c_p$ & $k$ & $e$\\ \hline
         \hline
         Ice & 916 & 2050 & 2.22 & 2041 \\
         Hexadecane & 833 & 1680 & 0.319 & 668.1\\
         \hline \addlinespace
    \end{tabular}
         
    \begin{tabular}{|c|c|c|c|c|}
        \multicolumn{5}{c}{Substrate} \\
             \hline
        - & $\rho$ & $c_p$ & $k$ & $e$\\ \hline
         \hline
         Aluminium & 2700 & 921 & 232 & 24019 \\
         \hline
         Glass & 2500 & 870 & 1.06 & 1518 \\
         \hline
    \end{tabular}
    \caption{Thermal properties of water, hexadecane (liquid and solid phases) and solid aluminium. $T_f$ ($\degree C$) is the melting temperature, $\rho$ ($kg.m^{-3}$) is the density, $c_p$ ($J.K^{-1}.kg^{-1}$) is the heat capacity, $k$ ($W.m^{-1}.K^{-1}$) is the thermal conductivity, $e$ ($J.m^{-2}.K^{-1}.s^{-\frac{1}{2}}$) is the effusivity and $L$ ($J.kg^{-1}$) is the latent heat.}
    \label{tab:liquid_properties}
\end{table}

\begin{table}[t]
    \centering

\begin{tabular}{ |p{4cm}||p{2.2cm}|p{2.2cm}|p{2.2cm}|}
 \hline
 \multicolumn{4}{|c|}{Compound drops : List of liquid used} \\
 \hline
 \centering{Liquids} & $\eta \ (mPa.s)$ & $\gamma \ (mN.m^{-1})$ & $\rho \ (kg.m^{-3})$ \\
 Water                                 & 1 & 72 & 1000\\
 Water - Glycerol I               & 57 & 68 & 1118\\
 Water - Glycerol II                & 88 & 68 & 1118\\
 Water - AOT & 1 & 25 & 1000 \\
 Octane & 0.55 & 21 & 700\\
 Dodecane                              & 1.3 & 25 & 745\\
  Hexadecane                            & 3.45 & 27 &  767\\
  Mineral oil & 29 & 35 & 840 \\
  Rapeseed oil & 69 & 33 & 920 \\
 \hline
\end{tabular}

\caption{Properties of the different liquids used to produce compound drops. $\eta$ is the viscosity, $\gamma$ the interfacial tension with air and $\rho$ the liquid density. The $AOT$ concentration is $C_{\mathrm{AOT}}=20 \ mM$.}
 \label{table4: compound drops}
\end{table}

\FloatBarrier 

\section*{Data availability}
The data used in this paper are available from authors upon reasonable
request.

\section*{Author contributions}

B.G. carried out most of the experiments. The data
analysis was carried out by B.G. and M.A. with help from H.K. and D.B.. All authors contributed to the writing of the paper.

\section*{Competing interests}
The authors declare no competing interests.

\section*{Acknowledgment}

This work is supported by the European Research Council (ERC) under the European Union’s Horizon Europe research and innovation programme (Grant agreement No. 101142159).



\begin{appendices}


\section{Comparison with Laan et al. and fitted effective parameters}

In the main text, we examined quantitatively the maximum spreading diameter $D_{\mathrm{max}}$ of such isothermal compound drops upon impact. We build our main model by extending the \citet{abbot2026kinematic} model to work for the outer and inner drops without adding any fitting parameter. Here, we show that comparing with previously proposed scaling law could have been done but introduces additional fitting parameter, which is undesirable. We compare the maximum spreading diameter of compound drops with the scaling law proposed by \citet{laan_maximum_2014}, which considers inertial, viscous and capillary effects. They propose a simple functional form given by:

\begin{equation}
    \frac{D_{\mathrm{max}}}{D_0} \sim \Re^{1/5} f(P)
    \label{eq4: Dmax scaling law compound drop appendix}
\end{equation}

with $f(P) = \frac{\sqrt{P}}{A+\sqrt{P}}$, $D_0$ the initial drop diameter and $A=1.24$ \cite{laan_maximum_2014}. The variable $P$, known as the impact parameter, is defined as $P=\We\Re^{-2/5}$, where $\We$ and $\Re$ are the Weber and Reynolds numbers, respectively. 
For simple drop impacts, the Weber and Reynolds numbers are defined as $\Re=\rho V_0 D_0 / \eta$ and $\We=\rho {V_0}^2 D_0 / \gamma$. Here, we define the Weber and Reynolds numbers for both phases. 

For the outer drop, $\Re_{\mathrm{out}}$ and $\We_{\mathrm{out}}$ are defined as:

\begin{equation}
    \Re_{\mathrm{out}}=\frac{\rho_{\mathrm{eff}} V_0 D_{\mathrm{0,out}}}{\eta_{\mathrm{out}}} \quad \quad \quad \We_{\mathrm{out}}=\frac{\rho_{\mathrm{eff}} {V_0}^2 D_{\mathrm{0,out}}}{\gamma_{\mathrm{out}}}
    \label{eq: Re and We outer drop appendix}
\end{equation}

The viscosity and the surface tension are the same as those of the outer layer. The drop size corresponds to the compound drop size. However, the density is given by the effective density which takes into account the density of the core drop. The effective density is $\rho_{\mathrm{eff}}=\left( \alpha \rho_{\mathrm{in}} + \rho_{\mathrm{out}}\left( 1-\alpha\right)\right)$.

For the inner drop, we have $\Re_{\mathrm{in}}$ and $\We_{\mathrm{in}}$ defined as :

\begin{equation}
    \Re_{\mathrm{in}}= \frac{\rho_{\mathrm{in}} V_0 D_{\mathrm{0,in}}}{\eta_{\mathrm{in}}} 
    \quad \quad \quad 
    \We_{\mathrm{in}} = \frac{\rho_{\mathrm{in}} {V_0}^2 D_{\mathrm{0,in}}}{\max\left(\gamma_{\mathrm{in}}, \gamma_{\mathrm{out}} \right)}
    \label{eq: Re and We inner drop}
\end{equation}

\begin{figure}[!ht]
\centering
\includegraphics[width=\textwidth]{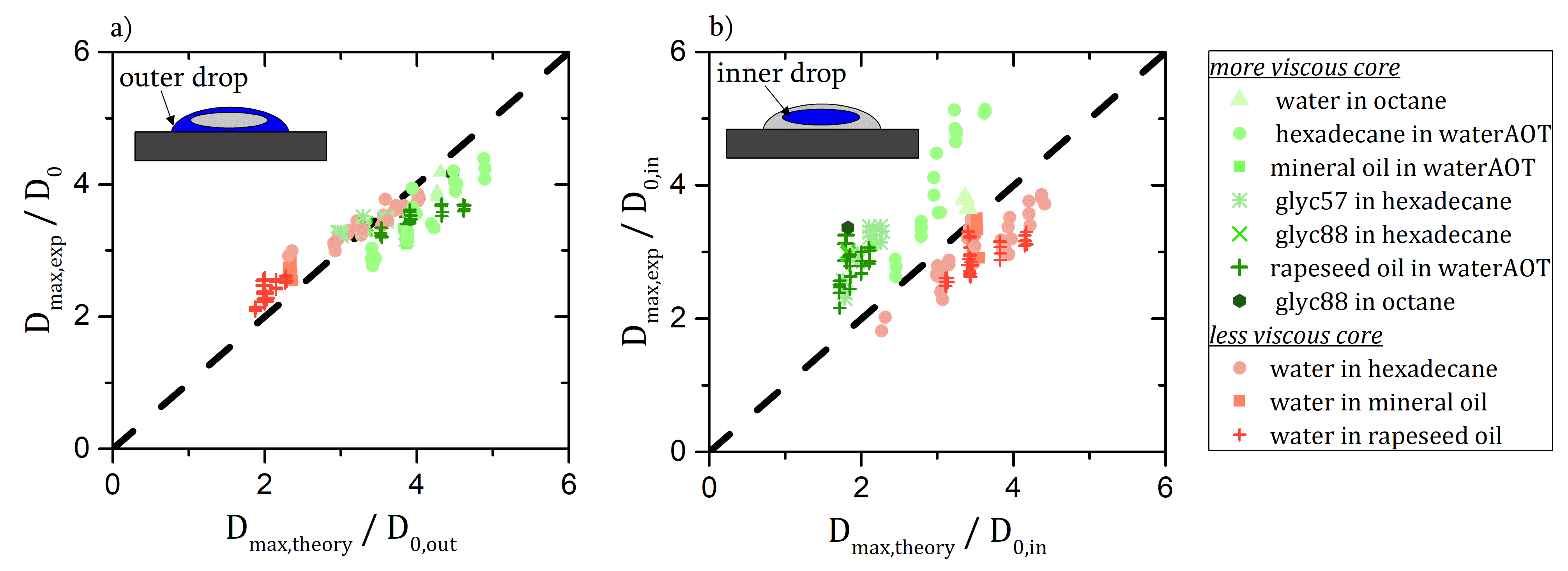}
    \caption{Comparison of the maximum spreading diameter of compound drops from the experiment with the scaling law of \citet{laan_maximum_2014} (equation \ref{eq4: Dmax scaling law compound drop appendix}). a) Outer (shell) drop and b) inner (core) drop are shown separately. The color scale illustrate the ratio of viscosity between the core and the shell drop.}
    \label{fig4: compound Dmax}
\end{figure}

The density and the viscosity are those of the inner liquid drop. The drop size corresponds to the inner drop size, given by $D_{\mathrm{0,in}}= \alpha^{1/3} D_{\mathrm{0,out}}$. The interfacial tension is chosen to be the maximum between the liquid-liquid interfacial tension $\gamma_{\mathrm{in}}$ \cite{wu_dynamic_1999,zeppieri_interfacial_2001} or the surface tension of the outer layer with air $\gamma_{\mathrm{out}}$. 
Different compound drop compositions are examined with different impact velocities and compared to the prediction of \citet{laan_maximum_2014}. The results are reported in Figure \ref{fig4: compound Dmax}.

In Figure \ref{fig4: compound Dmax} a, the maximum spreading diameter of the outer drop is compared to the prediction of equation \ref{eq4: Dmax scaling law compound drop appendix}. We note that the maximum spreading diameter of the outer shell is close to that of a simple drop: The presence of the core does not affect the spreading of the shell drop significantly.

In Figure \ref{fig4: compound Dmax} b, we report the maximum spreading diameter of the inner drop compared to the expected values. 
We note that the maximum spreading diameter of the core deviates from the expected diameter.
In the case of a core drop of smaller viscosity than the outer shell, the spreading of the core is slightly affected by the surrounding viscous shell. The maximum diameter is systematically below the expected diameter. The viscous dissipation is most probably increased due to the presence of the surrounding viscous shell which should reduce the spreading of the core drop, leading to a smaller maximum spreading diameter.
On the contrary, when the core drop is more viscous than the shell liquid, we note that the spreading of the core is enhanced by the presence of a less viscous shell drop, resulting in a higher maximum spreading diameter of the core: the core spreads more than what we could expect from the equivalent simple drop. The shell liquid acts as a lubricating fluid for the spreading of the more viscous core drop.

\vspace{2mm}

The maximum spreading diameter of both the core and the shell drop deviates from the prediction of equation \ref{eq4: Dmax scaling law compound drop appendix} and we found that the deviation depends on the difference of viscosity between the two liquid phases.
Based on these observations and as the model of equation \ref{eq4: Dmax scaling law compound drop appendix} predicts the maximum spreading diameter of simple drop impacts, we propose to introduce an effective viscosity $\eta_{\mathrm{eff}}$ such that equation \ref{eq4: Dmax scaling law compound drop appendix} becomes:

\begin{equation}
    \frac{D_{\mathrm{max}}}{D_0}\sim {\Re_{\mathrm{eff}}}^{1/5} * f\left(P_{\mathrm{eff}}\right)
    \label{eq:Laan visco_eff}
\end{equation}

with $P=\We\Re_{\mathrm{eff}}^{-2/5}$ and $\Re_{\mathrm{eff}}=\frac{\rho V_0 D_0}{\eta_{\mathrm{eff}}}$.
The effective viscosity $\eta_{\mathrm{eff}}$ is chosen as a free parameter and is adjusted in such a way that the maximum measured diameter of the compound drops (both the inner and outer drops) corresponds to the expected one from an effective equation~\ref{eq:Laan visco_eff}. 

\begin{figure}[!ht]
    \centering
    \includegraphics[width=0.90\textwidth]{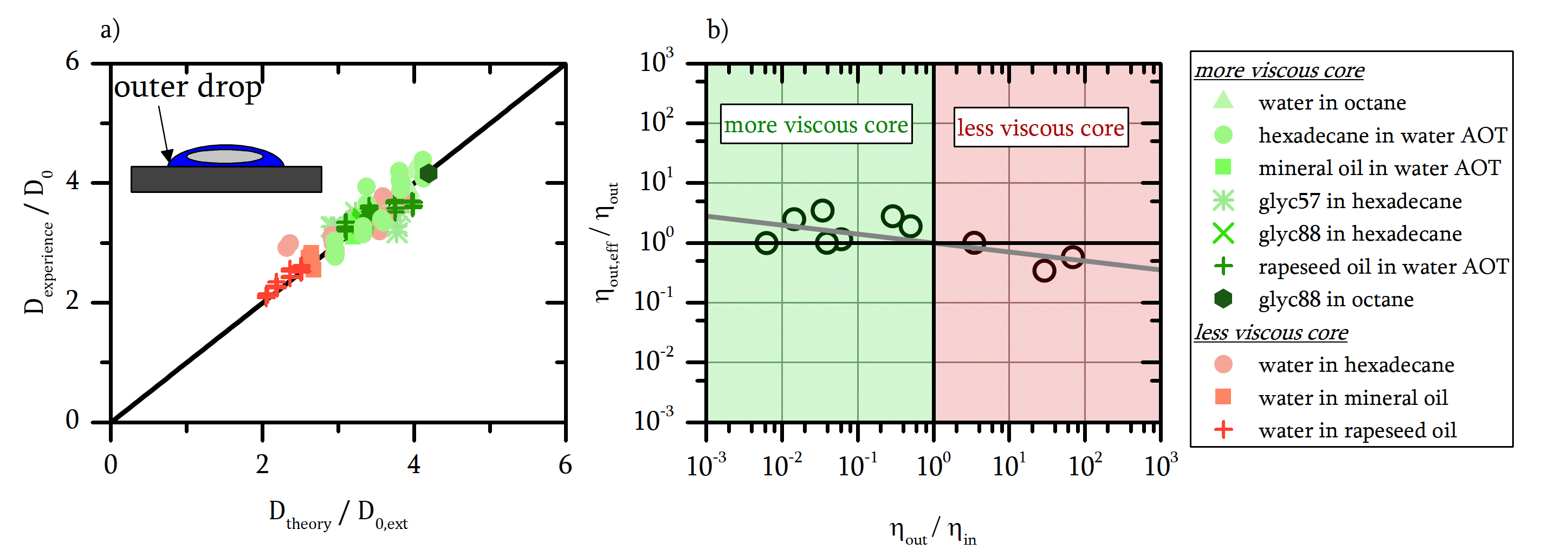}
    \caption{a) Maximum spreading diameter of the outer drop compared to the prediction of equation \ref{eq:Laan visco_eff} using the effective Reynolds and Weber number (see equation \ref{eq: Re and We outer drop appendix}). b) Normalized effective viscosity of the outer drop as a function of the viscosity ratio of the two phases. Power law of equation \ref{eq:power law outer} is represented as a gray line.}
    \label{fig:outer drop eta_eff}
\end{figure}

In Figure \ref{fig:outer drop eta_eff}a, we compare the maximum spreading diameter of the shell drop with the model of equation \ref{eq:Laan visco_eff} and along with the comparison, we report on Figure \ref{fig:outer drop eta_eff}b the values of the effective viscosity of the shell drop $\eta_{eff,out}$. The deviation with the model of simple drops is relatively small, as we have seen in Figure \ref{fig4: compound Dmax}a. Therefore, the effective viscosity needed is close to the viscosity of the liquid shell drop. Nevertheless, we find that the effective viscosity systematically depends on the ratio of viscosity between the core and the shell drop $\eta_{\mathrm{in}}/\eta_{\mathrm{out}}$ and approximately follows a power law as:

\begin{equation}
    \frac{\eta_{out,eff}}{\eta_{\mathrm{out}}} \approx \left(\frac{\eta_{\mathrm{in}}}{\eta_{\mathrm{out}}}\right)^{0.15}
    \label{eq:power law outer}
\end{equation}

\vspace{5mm}

The effective viscosity of the outer phase decreases with the ratio of outer to inner viscosity: the thinner the inner phase the larger the maximum diameter. 

\begin{figure}[!ht]
    \centering
    \includegraphics[width=0.90\textwidth]{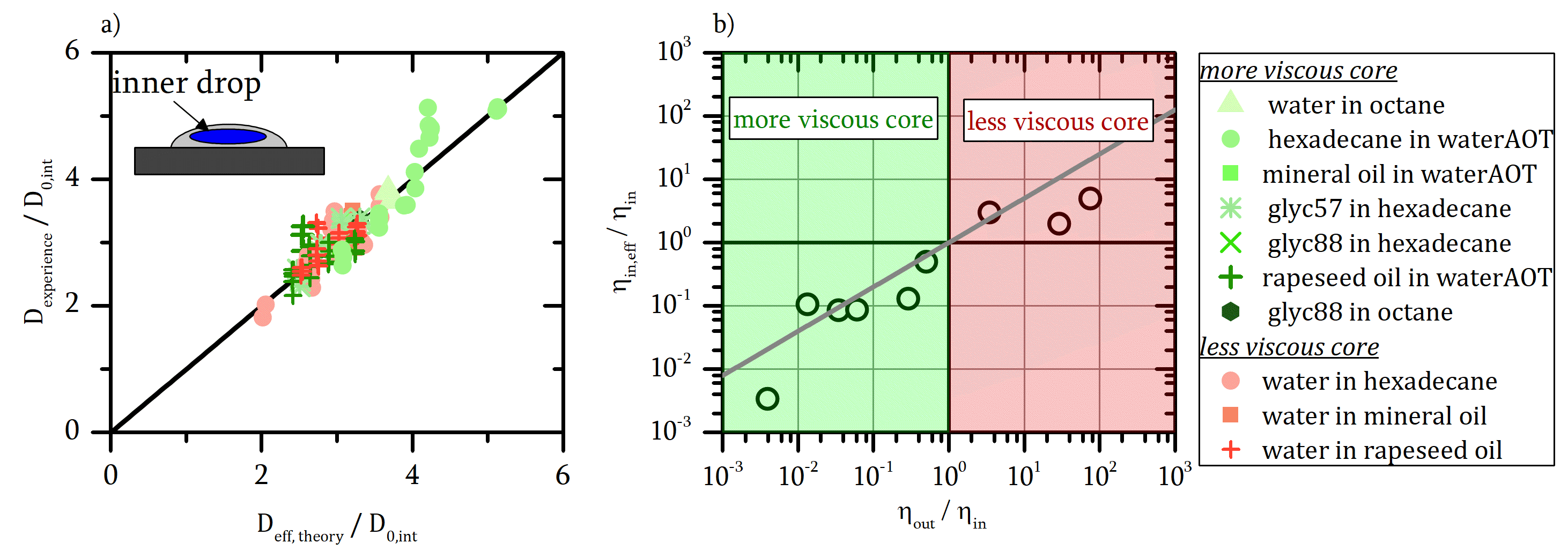}
    \caption{a) Maximum spreading diameter of the inner drop compared to the prediction of equation \ref{eq:Laan visco_eff} using the effective Reynolds and Weber number (see equation \ref{eq: Re and We inner drop}. b) Normalized effective viscosity of the inner drop as a function of the viscosity ratio of the two phases. Power law of equation \ref{eq:power law inner} is represented as a gray line.}
    \label{fig:inner drop eta_eff}
\end{figure}

For the inner drop, the comparison of the maximum spreading diameter with the model of equation \ref{eq:Laan visco_eff} is shown in Figure \ref{fig:inner drop eta_eff} along with the effective viscosity of the inner drop represented in Figure \ref{fig:inner drop eta_eff}.
As expected from the previous observations of Figure \ref{fig4: compound Dmax}b, we found that for a core drop with a lower  viscosity shell (e.g. $\eta_{\mathrm{out}} / \eta_{\mathrm{in}} < 1$), the spreading diameter is relatively larger and thus the effective viscosity needed is lower than the viscosity of the core drop. On the contrary, when the core liquid is less viscous than the shell fluid, the effective viscosity is higher than the real viscosity. The effective viscosity $\eta_{\mathrm{eff}}$ of the core drop strongly depends on the ratio of viscosities $\eta_{\mathrm{out}}/\eta_{\mathrm{in}}$. We find that the effective viscosity follows a power law of the form:

\begin{equation}
    \frac{\eta_{in,eff}}{\eta_{\mathrm{in}}} \approx \left(\frac{\eta_{\mathrm{out}}}{\eta_{\mathrm{in}}}\right)^{0.7}
    \label{eq:power law inner}
\end{equation}

Thus, the spreading of compound drops and the maximum spreading diameter in particular can be approximated using a power law effective viscosity whether for the outer shell or the core drop.


\end{appendices}

\section*{Supplementary Information}

Movie S1: Impact at $V_0=1.85m.s^{-1}$ of a water in hexadecane oil compound drop on an aluminium surface at $T=10.5 \degree C$.

Movie S2: Impact at $V_0=2.0m.s^{-1}$ of a hexadecane oil-in-water with AOT surfactant compound drop on an aluminium surface at $T=8.4 \degree C$.

\bibliography{mybib}

\end{document}